\documentclass[aip,reprint,jcp]{revtex4-1}

\usepackage{subfiles}
\usepackage{amsmath}
\newcommand*{\dif}{\mathop{}\mathrm{d}}

\usepackage{graphicx}
\graphicspath{{figures/}}

\usepackage{siunitx}
\DeclareSIUnit\atomicmassunit{u}
\DeclareSIUnit\angstrom{\text{Å}}

\usepackage{braket}

\newcommand{\hide}[1]{}

\usepackage{etoolbox}
\newtoggle{review}

\usepackage[normalem]{ulem}
\newcommand{\highlightred}[1]{\iftoggle{review}{\textcolor{red}{#1}}{#1}}
\newcommand{\strike}[1]{\iftoggle{review}{\textcolor{blue}{\sout{#1}}}{}}

\begin{document}

 
\title{Efficient implementation and performance analysis of the independent electron surface hopping method for dynamics at metal surfaces}



\author{James Gardner}
\author{Daniel Corken}
\affiliation{Department of Chemistry, University of Warwick, Gibbet Hill Road, Coventry CV4 7AL, United Kingdom}

\author{Svenja M. Janke}
\affiliation{Department of Chemistry, University of Warwick, Gibbet Hill Road, Coventry CV4 7AL, United Kingdom}
\affiliation{Institute of Advanced Study, University of Warwick, Coventry, UK}

\author{Scott Habershon}

\affiliation{Department of Chemistry, University of Warwick, Gibbet Hill Road, Coventry CV4 7AL, United Kingdom}

\author{Reinhard J. Maurer}
\affiliation{Department of Chemistry, University of Warwick, Gibbet Hill Road, Coventry CV4 7AL, United Kingdom}
\affiliation{Department of Physics, University of Warwick, Gibbet Hill Road, Coventry CV4 7AL, United Kingdom}
\email{r.maurer@warwick.ac.uk}


\date{\today}

\begin{abstract}
Independent electron surface hopping (IESH) is a computational algorithm for simulating the mixed quantum-classical molecular dynamics of adsorbate atoms and molecules interacting with metal surfaces. It is capable of modelling the nonadiabatic effects of electron-hole pair excitations on molecular dynamics.
Here we present a transparent, reliable, and efficient implementation of IESH, demonstrating its ability to predict scattering and desorption probabilities across a variety of systems, ranging from model Hamiltonians to full dimensional atomistic systems.
We further show how the algorithm can be modified to account for the application of an external bias potential, comparing its accuracy to results obtained using the hierarchical quantum master equation. Our results show that IESH is a practical method for modelling coupled electron-nuclear dynamics at metal surfaces, especially for highly energetic scattering events.
\end{abstract}


\maketitle

\section{Introduction}

When molecules interact with metal surfaces, nonadiabatic effects can strongly influence the dynamical reaction outcome. These effects arise from the excitation of electron-hole-pair (EHP) excitations.
\cite{huangVibrationalPromotionElectron2000, whiteConversionLargeamplitudeVibration2005}
As a consequence of their reliance on the Born--Oppenheimer approximation, classical adiabatic molecular dynamics methods for investigating the motion atomic and molecular adsorbates fail to accurately describe nonadiabatic effects.
To go beyond the Born--Oppenheimer approximation a variety of methods have been proposed.
A popular family of methods are based on trajectory surface hopping (TSH),
\cite{tullyMolecularDynamicsElectronic1990,shenviNonadiabaticDynamicsMetal2009,shenviDynamicalSteeringElectronic2009,shenviNonadiabaticDynamicsMetal2012,ouyangSurfaceHoppingManifold2015,douSurfaceHoppingManifold2015,douSurfaceHoppingManifold2015a,douBroadenedClassicalMaster2016,miaoComparisonSurfaceHopping2019,jinNonadiabaticDynamicsMetal2021,pradhanDetailedBalanceIndependent2022}
where the molecule moves according to one electronic state at a time, but sudden transitions between electronic states can occur during the dynamics if states are strongly coupled. Alternative approaches include mean-field approximations to the dynamics such as Ehrenfest dynamics
\cite{mclachlanVariationalSolutionTimedependent1964,grotemeyerElectronicEnergyDissipation2014,kirranderEhrenfestMethodsElectron2020,choiHighorderGeometricIntegrators2021}
or the molecular dynamics with electronic friction (MDEF) method.\cite{liNonadiabaticEffectsHydrogen1992,liMoleculardynamicsSimulationHydrogen1992,head-gordonMolecularDynamicsElectronic1995,douFrictionalEffectsMetal2015,novkoInitioMolecularDynamics2015,askerkaRoleTensorialElectronic2016,maurerInitioTensorialElectronic2016,boxDeterminingEffectHot2021} In the latter, nonadiabatic effects are captured by Langevin dynamics with system-bath interaction defined by the coupling to EHP excitations in the metal.
MDEF is easily applied to atomistic systems but has clear limitations.
Its assumption of weak \highlightred{nonadiabatic} coupling \strike{in the constant-coupling regime between the molecule and EHP excitations} prevents an accurate description of charge-transfer processes.\cite{ouyangDynamicsBarrierCrossings2016,boxDeterminingEffectHot2021}

Amongst the TSH methods for dynamics at metal surfaces, the predominant choice is independent electron surface hopping (IESH).
First introduced by Shenvi, Roy and Tully in 2009,
\cite{shenviNonadiabaticDynamicsMetal2009,shenviDynamicalSteeringElectronic2009} IESH was used to model the scattering dynamics of NO on a Au(111) surface.\cite{royModelHamiltonianInteraction2009}
Later, the method was revisited and augmented by the addition of thermostats for the nuclear and electronic degrees of freedom, with the purpose of enhancing the rate of electronic relaxation.
\cite{shenviNonadiabaticDynamicsMetal2012,miaoComparisonSurfaceHopping2019} More recently, the method has been further modified with a decoherence correction to enhance its ability to satisfy detailed balance. \cite{pradhanDetailedBalanceIndependent2022}

The IESH method has been shown to qualitatively capture certain experimental phenomena, such as nonadiabatic vibrational de-excitation during the state-to-state scattering of NO on Au(111),\cite{shenviNonadiabaticDynamicsMetal2009,shenviDynamicalSteeringElectronic2009} though some disagreements with experiment remained, which are likely due to shortcomings of the \highlightred{potential energy surfaces (PESs)} used for the simulations. \cite{krugerNOVibrationalEnergy2015,yinStrongVibrationalRelaxation2019,boxDeterminingEffectHot2021}
So far, application of IESH to realistic atomistic systems has remained limited to scattering of NO on Au(111),
likely due to the difficulty of obtaining the excited state PESs necessary to parametrize the model Hamiltonian, although recently there has been progress in this direction.\cite{mengPragmaticProtocolDetermining2022}
The increasing popularity of machine learning techniques may prove beneficial in this area, similar to recent progress accelerating MDEF simulations. \cite{zhangHotelectronEffectsReactive2019,maurerHotElectronEffects2019}

In this article, we present a reliable and efficient implementation of IESH, available in the recently-announced open-source NQCDynamics.jl package,\cite{gardnerNQCDynamicsJlJulia2022}
and demonstrate its application to a variety of systems.
We provide a detailed description of the implementation and present solutions to challenges encountered along the way.
These include optimization of the algorithm for computational efficiency, modelling the application of an external bias potential,
and initialization of the electronic subsystem in a given diabatic state.
In addition to the previously explored NO on Au(111) Hamiltonian model proposed by Roy, Shenvi and Tully\cite{royModelHamiltonianInteraction2009,shenviDynamicalSteeringElectronic2009}
and double-well electron transfer models,\cite{miaoComparisonSurfaceHopping2019}
we employ IESH to describe a 1D molecular desorption model, for which benchmark results based on the hierarchical quantum master equation method (HQME) exist.\cite{erpenbeckCurrentInducedBondRupture2018}
Further, we extend the same model to explore inelastic energy loss during scattering and compare the outcomes of IESH and MDEF.
By including a variety of models, both low-dimensional and atomistic, we highlight the transferability and robustness of our implementation.

We present an open-source implementation of the IESH algorithm that is transferable across different systems. This is facilitated by the NQCModels.jl framework previously introduced as part of NQCDynamics.jl.\cite{gardnerNQCDynamicsJlJulia2022}
Although a number of popular software solutions exist for TSH methods,\cite{Richter2011JCTC,barbattiNewtonXSurfacehoppingProgram2014,barbattiNewtonXPlatformNew2022} these do not currently treat metallic systems using IESH. When benchmarking IESH with one dimensional models, it has been previously compared to Marcus theory\cite{miaoComparisonSurfaceHopping2019,pradhanDetailedBalanceIndependent2022} but not a numerically exact method such as the HQME.\cite{schinabeckHierarchicalQuantumMaster2016}
Our efforts take a step to rectify this shortcoming by comparing to the benchmark data presented in Ref.~\citenum{erpenbeckCurrentInducedBondRupture2018}.

The remainder of the article is structured as follows.
In Sec.~\ref{sec:theory}, the IESH method and the Newns--Anderson Hamiltonian used to represent the coupled electron-nuclear system are introduced.
Here we discuss the specifics of the implementation and its adaptation to achieve the best possible results.
In Secs.~\ref{sec:NOAu} and~\ref{sec:double-well} we apply the IESH method to two previously explored models and demonstrate our ability to reproduce previous work.
Secs.~\ref{sec:desorption} and~\ref{sec:scattering} use the model of \citet{erpenbeckCurrentInducedBondRupture2018} to benchmark the performance of IESH for desorption and scattering processes.
For the former, we compare to the HQME and for the latter, MDEF.
We conclude in Sec.~\ref{sec:conclusion}, and discuss avenues for future development to enhance the utility of IESH for the description of nonadiabatic dynamics in condensed phase systems.

\section{Theory}\label{sec:theory}

\subsection{Newns-Anderson Hamiltonian}\label{sec:newns-anderson}

Before introducing the IESH method, it is useful to first review the Newns--Anderson (NA) Hamiltonian
\cite{andersonLocalizedMagneticStates1961,newnsSelfConsistentModelHydrogen1969}
(sometimes called Anderson--Holstein Hamiltonian\cite{douNonadiabaticMolecularDynamics2020}).
This model forms the starting point for the IESH method, and the system under investigation must be mapped into the NA form before IESH can be applied.

The NA Hamiltonian $\hat{H}_\text{NA}$ contains a molecule with a single electronic state coupled to a bath of electronic states, and can be written as
\begin{equation}
    \hat{H}_\text{NA}(\hat{x},\hat{p}) = \frac{\hat{p}^2}{2m} + U_0(\hat{x}) + \hat{H}_\text{el}(\hat{x}),
    \label{eq:NA_hamiltonian}
\end{equation}
where the electronic Hamiltonian is 
\begin{equation}
    \hat{H}_\text{el}(\hat{x}) = h(\hat{x})\hat{d}^\dag \hat{d} + \int_a^b \dif\epsilon \, \epsilon \hat{c}_\epsilon^\dag \hat{c}_\epsilon
    + \int_a^b \dif\epsilon \, V(\epsilon; \hat{x}) (\hat{d}^\dag \hat{c}_\epsilon + \hat{c}_\epsilon^\dag \hat{d}).
    \label{eq:electronic_hamiltonian}
\end{equation}
For simplicity, all equations are presented for a single nuclear degree of freedom $\hat{x}$ with conjugate momentum $\hat{p}$ and mass $m$.
A circumflex is given to the variables that represent quantum mechanical operators.
The state-independent PES is given by $U_0(\hat{x})$.
Inside the electronic Hamiltonian $\hat{H}_\text{el}$, 
$\hat{d}^\dag (\hat{d})$ are the creation (annihilation) operators for an electron in the molecular state and
$\hat{c}_\epsilon^\dag (\hat{c}_\epsilon)$ are the creation (annihilation) operators for an electron in the metal with energy $\epsilon$.
When the molecular state is occupied, $h(\hat{x}) = U_1(\hat{x}) - U_0(\hat{x})$ is added as a further contribution to the system potential energy.
The second term in Eq.~\ref{eq:electronic_hamiltonian} describes a band of non-interacting electrons that models the metallic continuum.
The range of energies is defined by the band edges $a$ and $b$.
The final term facilitates the coupling between the molecular state and the metal where the magnitude of the coupling is determined by the function $V(\epsilon; \hat{x})$.
The hybridization function that characterizes the molecule-metal coupling is\cite{devegaHowDiscretizeQuantum2015}
\begin{equation}
    \Gamma(\epsilon; \hat{x}) = 2\pi \int_a^b \dif\epsilon' |V(\epsilon'; \hat{x})|^2 \delta(\epsilon - \epsilon') = 2\pi |V(\epsilon; \hat{x})|^2.
    \label{eq:hybridization}
\end{equation}
However, for each of the models investigated in this article,
the hybridization function is always independent of energy such that $\Gamma(\epsilon; \hat{x}) \to \Gamma(\hat{x})$ and $V(\epsilon; \hat{x}) \to V(\hat{x})$.

To consider electronic transitions between a manageable finite number of states in surface hopping, it is necessary to discretize the continuum of electronic states.
\cite{shenviEfficientDiscretizationContinuum2008,devegaHowDiscretizeQuantum2015}
Usually, this is done by using a trapezoid rule to generate evenly spaced states,\cite{miaoComparisonSurfaceHopping2019}
or Gaussian quadrature to generate a set of states that are unevenly distributed.
\cite{shenviNonadiabaticDynamicsMetal2009}
The latter method has the advantage of increasing the number of states at the Fermi level, which might lead to faster convergence with respect to the number of states for some methods, including IESH.

Upon discretization, the bath and coupling undergo the following transformations:
\begin{align}
\int_a^b \dif\epsilon \, \epsilon \hat{c}_\epsilon^\dag \hat{c}_\epsilon
&\to \sum_{k=1}^N \epsilon_k \hat{c}_k^\dag \hat{c}_k
\label{eq:bath_discrete}
\\
\int_a^b \dif\epsilon \, V(\epsilon; \hat{x}) (\hat{d}^\dag \hat{c}_\epsilon + \hat{c}_\epsilon^\dag \hat{d})
&\to \sum_{k=1}^N V_k(\hat{x}) (\hat{d}^\dag \hat{c}_k +  \hat{c}_k^\dag \hat{d}),
\label{eq:coupling_discrete}
\end{align}
where the integrals have been replaced by sums and the set of $\{\epsilon_k\}$ span the width of the band (from $a$ to $b$).
The specific values for $\epsilon_k$ and $V_k$ depend upon the choice of discretization strategy where $N$ is the total number of states.
The reader is referred to Appendix~\ref{sec:discretization} for a description of the two most popular discretizations.
Throughout Sec.~\ref{sec:results} we use both discretization methods and determine which one is most effective.
Upon discretization, the hybridization function in Eq.~\ref{eq:hybridization} becomes
\begin{equation}
    \Gamma(\epsilon; \hat{x}) = \highlightred{2\pi}\sum_{k=1}^N \highlightred{|V(\epsilon;\hat{x})|^2}\delta(\epsilon - \epsilon_k).
    \label{eq:discrete_hybridization}
\end{equation}

Using the transformations in Eqs.~\ref{eq:bath_discrete} and~\ref{eq:coupling_discrete} the electronic Hamiltonian becomes
\begin{equation}
    \hat{H}_\text{el}^{N}(\hat{x}) = h(\hat{x})\hat{d}^\dag \hat{d}
    + \sum_{k=1}^N \epsilon_k \hat{c}_k^\dag \hat{c}_k
    + \sum_{k=1}^N V_k(\hat{x}) (\hat{d}^\dag \hat{c}_k + \hat{c}_k^\dag \hat{d}).
    \label{eq:discrete_electronic_hamiltonian}
\end{equation}
Combined with Eq.~\ref{eq:NA_hamiltonian}, Eq.~\ref{eq:discrete_electronic_hamiltonian} provides the Hamiltonian foundation for IESH.

\subsection{Independent electron surface hopping}\label{sec:iesh}

The IESH method is based upon an extension of traditional fewest-switches surface hopping\cite{tullyMolecularDynamicsElectronic1990}
to a metal system with many electrons.
Due to the assumption of independent electrons, it is possible to dramatically reduce the complexity of the many-electron Hamiltonian.
In IESH, the nuclei follow the equations of motion generated by the time-dependent classical Hamiltonian
\begin{equation}
    H_\text{IESH}(x,p,t) = \frac{p^2}{2m} + U_0(x) + \sum_{k \in \mathbf{s}(t)} \lambda_k(x).
    \label{eq:classical_hamiltonian}
\end{equation}
Note that the nuclear variables no longer have a circumflex as the quantum nuclei are being approximated by classical particles.
The potential energy consists of two components: the state-independent contribution $U_0(x)$,
and the energy of each occupied single-electron state $\lambda_k(x)$.
In the electronic Hamiltonian (Eq.~\ref{eq:discrete_electronic_hamiltonian}) the electrons are non-interacting,
so the eigenvalues $\{\lambda_k(x)\}$ of the corresponding single-electron Hamiltonian can be used to obtain the total energy.
The time-dependence is introduced by the occupation vector $\mathbf{s}(t)$ that contains the indices of the occupied states at time $t$,
constraining the summation such that the energy receives contributions from only occupied states.

\newcommand{\ketbra}[2]{\ket{#1}\!\bra{#2}}

Alongside the classical nuclear dynamics generated by Eq.~\ref{eq:classical_hamiltonian},
the wave function of each electron is propagated by the single-electron Hamiltonian
\begin{equation}
    \hat{H}_\text{el}^1(x) = h(x)\ketbra{d}{d}
    + \sum_{k=1}^N \epsilon_k \ketbra{k}{k}
    + \sum_{k=1}^N V_k(x) (\ketbra{k}{d} + \ketbra{d}{k}),
    \label{eq:single_electron_discrete_electronic_hamiltonian}
\end{equation}
where $\ket{k}$ and $\ket{d}$ correspond to the single electron states acted upon by the operators $\hat{c}_k$ and $\hat{d}$, respectively.
In the adiabatic basis, the time-dependent Schr\"odinger equation for each electron becomes
\begin{equation}
    i\hbar \dot{c}_{\highlightred{k}} = \lambda_k(x) c_{\highlightred{k}} - i\hbar \sum_j \frac{p}{m} d_{jk}(x) c_{\highlightred{j}},
    \label{eq:electronic_schrodinger_equation}
\end{equation}
where $\{c_k\}$ are the complex expansion coefficients for each electron and
$d_{jk}$ is the nonadiabatic coupling between adiabatic states $k$ and $j$, given by
\begin{equation}
    d_{jk}(x) = \frac{\left(Q^\dag(x) \frac{\partial \hat{H}_\text{el}^1}{\partial x} Q(x) \right)_{jk}}{\lambda_j - \lambda_k},
\end{equation}
where $Q(x)$ is the transformation matrix that converts from the diabatic to the adiabatic representation,
containing the eigenvectors of $\hat{H}_\text{el}^1$.

Eqs.~\ref{eq:classical_hamiltonian} and~\ref{eq:electronic_schrodinger_equation} describe the motion of the nuclear and electronic subsystems.
The coupling between the subsystems is facilitated by surface hopping, where $\mathbf{s}(t)$ changes due to the electronic dynamics,
leading to a change in the PES in Eq.~\ref{eq:classical_hamiltonian}.
In IESH only single electron hops may occur during each time step $\Delta t$, equivalent to changing one element of $\mathbf{s}(t)$.
To determine whether a hop should occur, the probability for each electron to hop to each unoccupied state is calculated.
For an electron occupying state $k$, the probability to hop to state $j$ during each time step is given by
\begin{equation}
    g_{k \to j} = \max\left(\frac{B_{jk}\Delta t}{A_{kk}}, 0\right),
    \label{eq:hopping_probability}
\end{equation}
with
\begin{equation}
    B_{jk} = -2 \text{Re}(A_{kj}^*) \frac{p}{m} d_{jk}.
    \label{eq:hopping_rate}
\end{equation}
Eqs.~\ref{eq:hopping_probability} and~\ref{eq:hopping_rate} are equivalent to their original fewest-switches counterparts.\cite{tullyMolecularDynamicsElectronic1990}
The key difference lies in the evaluation of the matrix elements $A_{kj}$ and $A_{kk}$ of the full electronic density matrix $\ketbra{\psi}{\psi}$,
\begin{equation}
    A_{kj} = \braket{\mathbf{k}|\psi}\braket{\psi|\mathbf{j}},
\end{equation}
where each inner product is calculated from the determinant of the overlap matrix $S$ between the discrete electronic occupation vector and the single electron wave functions:
\begin{equation}
    \braket{\mathbf{k}|\psi} = \det(S),
\end{equation}
with the elements of the overlap matrix given by
\begin{equation}
    S_{ij} = c_{k_{i}}^{(j)},
\end{equation}
where $k_i$ is the state occupied by electron $i$ in the occupation vector $\mathbf{k}$,
and $\mathbf{c}^{(j)}$ is the vector of wavefunction coefficients for electron $j$.

After the hopping probabilities have been calculated, the algorithm proceeds exactly as in the standard fewest-switches algorithm.\cite{tullyMolecularDynamicsElectronic1990,shenviNonadiabaticDynamicsMetal2009}
A possible hop is selected by sampling the hopping probabilities, comparing to a uniform random number between 0 and 1.
If a hop is selected the velocity is rescaled in the direction of the nonadiabatic coupling to ensure the total energy is conserved.
If there is sufficient kinetic energy to successfully rescale the velocity, then the hop proceeds and the state vector $\mathbf{s}(t + \Delta t)$ is modified.
In the case of insufficient kinetic energy, the hop is rejected and the state remains unchanged.

Before proceeding to present our implementation, it is useful to discuss the electronic chemical potential in the context of IESH.
This will become particularly relevant in the simulations featured in Sec.~\ref{sec:desorption}.
In IESH, the number of electrons is chosen such that the lowest energy configuration fills all the states up to the desired chemical potential.
This allows us to explicitly specify the chemical potential of the metal in our system.
In previous work, the chemical potential has always been set equal to 0,\cite{shenviDynamicalSteeringElectronic2009,shenviNonadiabaticDynamicsMetal2009,shenviNonadiabaticDynamicsMetal2012,miaoComparisonSurfaceHopping2019}
though the number of electrons can be easily modified to change the chemical potential.
However, it is important to consider that the Gaussian quadrature discretization method was designed to increase the density of states at the Fermi level to improve the convergence of IESH.
By modifying the chemical potential, the region containing more finely spaced electronic states no longer aligns with the Fermi level of the metal and more states may be needed to converge the simulation.
To regain the benefits of the Gaussian quadrature method, the discretization can be modified to ensure the region of highest density remains close to the Fermi level.
For simulations in the wide band limit, where the band edges are treated as a convergence parameter, it is possible to shift the entire band to ensure the center of the band aligns with the Fermi level.
Where the band width is physically motivated, such as in the choice of \SI{7}{\eV} for Au,\cite{shenviNonadiabaticDynamicsMetal2009}
it is instead necessary to modify the limits of the integrals in Eq.~\ref{eq:gaussian-quadrature}.

\subsection{Implementation details}

We have implemented the IESH algorithm in the open-source software package NQCDynamics.jl. \cite{gardnerNQCDynamicsJlJulia2022}
Since IESH demonstrates algorithmic similarity with other TSH methods, for example like fewest-switches surface hopping it also requires velocity rescaling upon a successful hop, it has been implemented by reusing and extending the existing TSH  functionality.
The algorithm proceeds via numerical integration of the coupled differential equations of motion for the nuclear and electronic subsystems,
along with execution of a callback function at every time step that performs the surface hopping procedure.

The dynamics are propagated using the DifferentialEquations.jl software suite\cite{rackauckasDifferentialEquationsJlPerformant2017}
where it is possible to choose from a variety of solvers for general differential equations.
However, for IESH we were able to achieve better performance by implementing a custom integration algorithm,
specifically tailored to the structure of the IESH equations of motion.
We implement the integration using the velocity Verlet algorithm\cite{leimkuhlerSimulatingHamiltonianDynamics2005} for the nuclei,
coupled to a linear exponential integrator for the Schr\"odinger equation.

Our implementation of the electronic integrator is as follows.
First, Eq.~\ref{eq:electronic_schrodinger_equation} is rewritten in matrix form as
\begin{equation}
    i\hbar \dot{\mathbf{c}} = \mathbf{A} \mathbf{c},
\end{equation}
with
\begin{equation}
    A_{jk} = \begin{cases}
        \lambda_j                   & j = k\\
        -i\hbar\frac{p}{m}d_{jk}(x) & j \neq k\\
    \end{cases}.
\end{equation}
The solution is given by
\begin{equation}
    \mathbf{c}(t + \Delta t) = \exp\left(- \frac{i}{\hbar} \mathbf{A} \Delta t\right)\mathbf{c}(t),
    \label{eq:exponential_integrator}
\end{equation}
which is exact for any $\Delta t$ when the nuclei are frozen.
Application of Eq.~\ref{eq:exponential_integrator} requires that the matrix exponential be computed.
Given that $\mathbf{A}$ is hermitian, this can be done efficiently via diagonalization $\mathbf{D} = \mathbf{P}^\dag \mathbf{A} \mathbf{P}$
so that Eq.~\ref{eq:exponential_integrator} becomes
\begin{equation}
    \mathbf{c}(t + \Delta t) = \mathbf{P} \exp\left(- \frac{i}{\hbar} \mathbf{D} \Delta t\right) \mathbf{P}^\dag \mathbf{c}(t).
    \label{eq:fast_exponential_integrator}
\end{equation}
The advantage of Eq.~\ref{eq:fast_exponential_integrator} over Eq.~\ref{eq:exponential_integrator} is that it can use
highly optimized BLAS and LAPACK routines to perform the matrix multiplication and eigendecomposition steps.


Although the exponential algorithm achieves the exact solution only when the nuclei are frozen, such that $\mathbf{A}$ is time-independent,
it can be expected to perform well when the nonadiabatic coupling changes only a small amount during each time step. 
To circumvent this limitation, it is possible to use a smaller time step for the electronic integrator,
where the nonadiabatic couplings, velocities and eigenvalues are interpolated between nuclear time steps.
Interpolation techniques of this variety are popular in many TSH software packages. 
\cite{barbattiNewtonXSurfacehoppingProgram2014,maiNonadiabaticDynamicsSHARC2018}
For IESH, the same technique can be applied.
However, the size of the nuclear time step in IESH should be curtailed by the magnitude of the hopping probabilities.
As noted previously,\cite{shenviNonadiabaticDynamicsMetal2009} care must be taken to ensure that the probability during each individual step remains small,
otherwise multiple hops per time step would be required to recover the correct electronic dynamics.
We found that the hopping limitation on the nuclear time step became dominant well before the quality of the electronic integration begins to degrade.
For IESH dynamics to benefit from sub-stepping for the electronic integration, it would first be necessary to modify the hopping scheme to allow for larger nuclear time steps.

The IESH algorithm experiences unfavorable scaling with the number of electronic states included in the Hamiltonian.
In the case of $M$ electronic states and $M/2$ electrons, it is necessary to solve $M/2$ sets of $M$ equations,
each of which is coupled to the motion of the nuclei.
Further, to calculate the hopping probabilities, the determinant of the overlap matrix $S$ must be computed for all possible hops,
of which there are $O(M^2)$.
Combined, each of these components can become overwhelming, limiting the maximum number of states that can be used in the simulation.
This limitation is visible in previous work,
where the number of states used has rarely exceeded 100 and the number of trajectories is similarly restricted,
even for analytic models.
\cite{shenviNonadiabaticDynamicsMetal2009,douNonadiabaticMolecularDynamics2020}

One could argue that the performance limitations of IESH are secondary compared to the cost of obtaining the necessary electronic quantities using \textit{ab initio} methods.
However, the advent of machine learning methods in quantum chemistry has significantly enhanced the possibilities for performing efficient  energy and force evaluations for high-dimensional systems,
\cite{jiangHighFidelityPotentialEnergy2020,westermayrPerspectiveIntegratingMachine2021}
and it is likely that the computational complexity of the IESH algorithm can become prohibitive when the electronic evaluations become sufficiently fast.
A strategy that we have implemented to improve the performance is based upon minimizing the number of determinants evaluated at each step.

\highlightred{
The probability that any single hop will occur is given by
\begin{equation}
P_\text{max} = \sum_{j} g_{k \to j}
\end{equation}
where the summation gives the cumulative probability of each electron hopping to all unoccupied states $j$.
If $P_\text{max}$ is less than the generated random number, a hop cannot occur.
An upper bound for $P_\text{max}$ is given by
\begin{equation}
P_\text{max}^\text{est} =
\frac{-2\Delta t }{A_{kk}}
\left(
\left|\text{Re}\braket{k | \psi}\right|
+ \left|\text{Im}\braket{k | \psi}\right|
\right)
\sum_{j} \left|\frac{p}{m} d_{jk}\right|
\label{eq:hopping-estimate}
\end{equation}
which can be obtained by assuming $\left|\braket{j|\psi}\right|^2 = 1$ for all configurations $j$,
without the expensive calculation of all $\braket{j|\psi}$ elements.
By using the upper bound $P_\text{max}^\text{est}$ to reject hops in regions where the nonadiabatic couplings are small there can be significant performance improvements.
}

\strike{
For each pair of states, before calculating the determinant, we estimate the probability with
where $\text{Re}(A_{kj})$ in Eq.~\ref{eq:hopping_rate} has been replaced by 1 to provide an upper bound.
If the probability estimate is lower than the generated random number, it is sufficient to reject the hop immediately without the need to evaluate the expensive determinant.
In regions where the nonadiabatic couplings are small, this method can bring significant performance improvements.
}

\begin{figure}
    \includegraphics{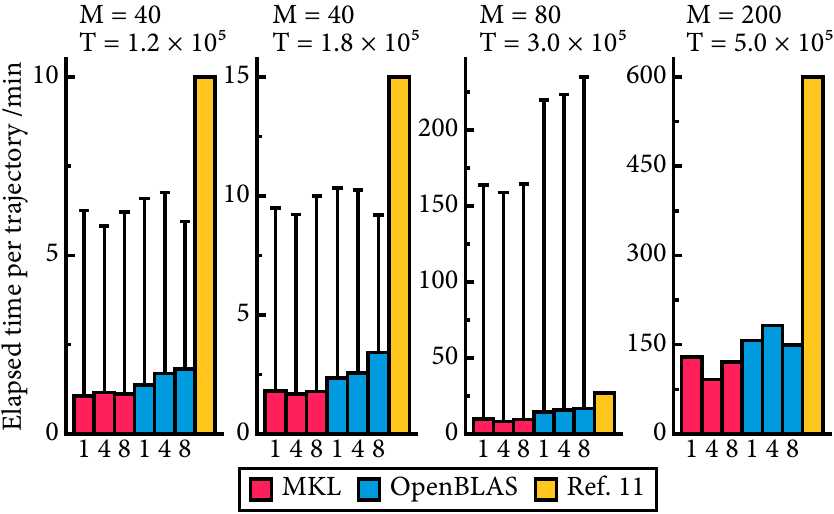}
    \caption{IESH performance of our implementation compared to the implementation used by \citet{miaoComparisonSurfaceHopping2019}.
    Table 1 of Ref.~\citenum{miaoComparisonSurfaceHopping2019} provides the reference timings to which we compare.
    For our results, the value of each bar is the minimum time to run a single trajectory from a sample of five trajectories.
    \highlightred{
    The error bars show the result when the hopping estimate in Eq.~\ref{eq:hopping-estimate} is switched off.
    }
    Results are presented for different choices of BLAS and LAPACK backends, comparing single-threaded and multi-threaded performance.
    For each backend, we used 1, 4 and 8 threads as labelled \highlightred{underneath} each bar.
    The simulations are the same as those described in Sec.~\ref{sec:double-well}, modified with the parameters shown above each chart, where
    $M$ is the number of bath states, and $T$ is the total number of time steps ($\Delta t = 10$).
    }
    \label{fig:performance}
\end{figure}

To evaluate the efficiency of our implementation, we have compared the time taken to run IESH simulations in a double well potential as described in Table 1 of Ref.~\citenum{miaoComparisonSurfaceHopping2019}.
Although we are not able to run our simulations on the same hardware, by comparing to the performance reported by Miao \emph{et al.}\cite{miaoComparisonSurfaceHopping2019} we obtain an approximate performance estimate.
Our results were obtained using Intel Xeon Gold 6248R \SI{3.0}{\giga\hertz} processors, Julia version 1.8.0, MKL 2022.2, and OpenBLAS 0.3.20.
Fig.~\ref{fig:performance} shows the elapsed real time to run simulations described in Sec.~\ref{sec:double-well}.
With the caveat that the simulation hardware is not the same, it appears that our implementation significantly outperforms the previously published result.
This trend is seen for both few and many bath states.

\highlightred{
The necessity of the hopping estimate procedure is clearly shown by the significant performance improvements
in Fig.~\ref{fig:performance}.
For $M=40$, we observe a fivefold reduction in runtime (height of error bars compared against blue and red bars), and greater than a tenfold reduction for $M=80$.
For $M=200$ the performance improvement is so great that a single trajectory without the hopping estimate does not finish within the walltime limit of 24 hours.
This result leads us to suspect that the implementation of Ref.~\citenum{miaoComparisonSurfaceHopping2019}
likely also includes a procedure to mitigate the poor scaling of the hopping probability calculations.
}

Furthermore, we note that the choice of BLAS and LAPACK provider can have significant effects on the simulation time.
Comparing MKL and OpenBLAS, we observe that MKL consistently achieves better performance, albeit possibly CPU architecture dependent. Given that we have used an Intel CPU, this result may not be surprising.
The superior performance of MKL over OpenBLAS has also been observed previously in the quantum chemistry package Fermi.jl.\cite{aroeiraFermiJlModern2022}
The multithreading behaviour also matches expectation.
When $M=40$, increasing the number of threads does not improve the performance for either MKL or OpenBLAS.
This is likely because the matrices are not sufficiently large and the threading overhead dominates.
However, on increasing to $M=80$ and $M=200$, the higher thread counts for MKL lead to a moderate speedup.
These results highlight the importance of optimizing the computational parameters for the particular system of interest.

\section{Results and discussion}\label{sec:results}
\subsection{Vibrational energy dissipation of NO on A\lowercase{u}(111)}\label{sec:NOAu}

Since the introduction of IESH, it has mostly been applied
\cite{shenviNonadiabaticDynamicsMetal2009,shenviDynamicalSteeringElectronic2009,shenviNonadiabaticDynamicsMetal2012,krugerNOVibrationalEnergy2015}
to an analytic NO on Au(111) model Hamiltonian parametrized against DFT data.\cite{royModelHamiltonianInteraction2009}
Although it has been shown that the model is incapable of describing experimental vibrational relaxation probabilities\cite{krugerNOVibrationalEnergy2015} as it was too soft allowing too facile energy transfer from translation to other degrees of freedom,\cite{yinStrongVibrationalRelaxation2019}
it remains a useful model to demonstrate the ability of our implementation to treat atomistic systems.
In this section, we examine the vibrational relaxation of an NO molecule bound to the Au(111) surface above the HCP site.

As described previously,\cite{shenviNonadiabaticDynamicsMetal2009} the parameters of the original two-state  model\cite{royModelHamiltonianInteraction2009}
must be modified such that the discretized NA Hamiltonian recovers the original ground state of the \textit{ab initio} calculations. This will be required for any parametrization that is based on an initial diabatic two-state model featuring a neutral molecular diabat and a charged excited-state diabat.
In Appendix~\ref{sec:mapping-hamiltonian} we provide a description of this modification procedure for a general two state \textit{ab initio} system.
However, to avoid reimplementing the analytic model and refitting the parameters, we were provided access to the Fortran implementation of the model developed by Roy et al.,\cite{royModelHamiltonianInteraction2009} which we couple to our code to directly access the energies and gradients using Julia's language interoperability features. \cite{bezansonJuliaFreshApproach2017}
For our simulations we retain the full set of refitted parameters present in the Fortran program, provided in Table~\ref{tab:noau-parameters}.
For this model, the only parameters that have changed significantly from their two-state values\cite{royModelHamiltonianInteraction2009} are those that  determine the magnitude of the coupling terms: $A_2$ and $B_2$.
To use the coupling values in the discretized expressions given in Appendix~\ref{sec:discretization}, both $A_2$ and $B_2$ are additionally multiplied by $\Delta E^{-\frac{1}{2}}$, where $\Delta E$ is the bandwidth of \SI{7}{\eV}.
Note that this transformation converts the units of the coupling element to $(\text{energy})^\frac{1}{2}$,
which upon insertion into the discretized expressions recovers the correct unit of energy.
This modification is essential to ensure that the NA Hamiltonian preserves the original ground state PES.

\begin{table}
    \caption{\label{tab:noau-parameters}
    Parameters for the two state diabatic Hamiltonian used to assemble the NO/Au(111) Newns-Anderson Hamiltonian.
    The functions that use these parameters are presented in Ref.~\cite{royModelHamiltonianInteraction2009}.
    The values that we present are taken from the Fortran implementation\cite{krugerNOVibrationalEnergy2015}
    where the parameters have been refitted to ensure the Newns-Anderson Hamiltonian preserves the original ground state PES.
    The only two parameters that differ significantly from their original values are $A_2$  and $B_2$.
    Each value is shown to a precision of five significant figures.
    }
    \begin{ruledtabular}
    \begin{tabular}{lll|lll}
    $A_0$ & \num{457000.05} & \si{\kilo\joule\per\mol}
    &
    $\alpha_0$ & \num{3.7526} & \si{\per\angstrom}
    \\
    $A_1$ & \num{457000.05} & \si{\kilo\joule\per\mol}
    &
    $\alpha_1$ & \num{3.7526} & \si{\per\angstrom}
    \\
    $A_2$& \num{16.749} & \si{\kilo\joule\per\mol}
    &
    $\beta_0$ & \num{2.9728} & \si{\per\angstrom}
    \\
    $A_3$& \num{0.0061715} & \num{1}
    &
    $\beta_1$ & \num{1.9101} & \si{\per\angstrom}
    \\
    $B_0$ & \num{30789} & \si{\kilo\joule\per\mol}
    &
    $\gamma_0$ & \num{2.7430} & \si{\per\angstrom}
    \\
    $B_1$ & \num{23.860} & \si{\kilo\joule\per\mol}
    &
    $\gamma_1$& \num{2.4709} & \si{\per\angstrom}
    \\
    $B_2$& \num{70.526} & \si{\kilo\joule\per\mol}
    &
    $\gamma_2$& \num{1.3535} & \si{\per\angstrom}
    \\
    $B_3$& \num{0.0047002} & \num{1}
    &
    $\gamma_3$& \num{1.9598} & \si{\per\angstrom}
    \\
    $C$ & \num{1.2558} & \si{\angstrom}
    &
    $r_0^\text{N-O}$ & \num{1.1508} & \si{\angstrom}
    \\
    $D$ & \num{347.22} & \si{\kilo\joule\per\mol\angstrom}
    &
    $r_1^{N-O}$& \num{1.2929} & \si{\angstrom}
    \\
    $E_a$& \num{-0.67540} & \si{\kilo\joule\per\mol}
    &
    $r_1^\text{Au-N}$ & \num{2.3896} & \si{\angstrom}
    \\
    $F_0$ & \num{638.50} & \si{\kilo\joule\per\mol}
    &
    $z_\text{image}$ & \num{1.1536} & \si{\angstrom}
    \\
    $F_1$& \num{495.98} & \si{\kilo\joule\per\mol}
    &
    $\phi$& \num{511.37} & \si{\kilo\joule\per\mol}
    \\
    \end{tabular}
    \end{ruledtabular}
\end{table}

\begin{figure}
\includegraphics{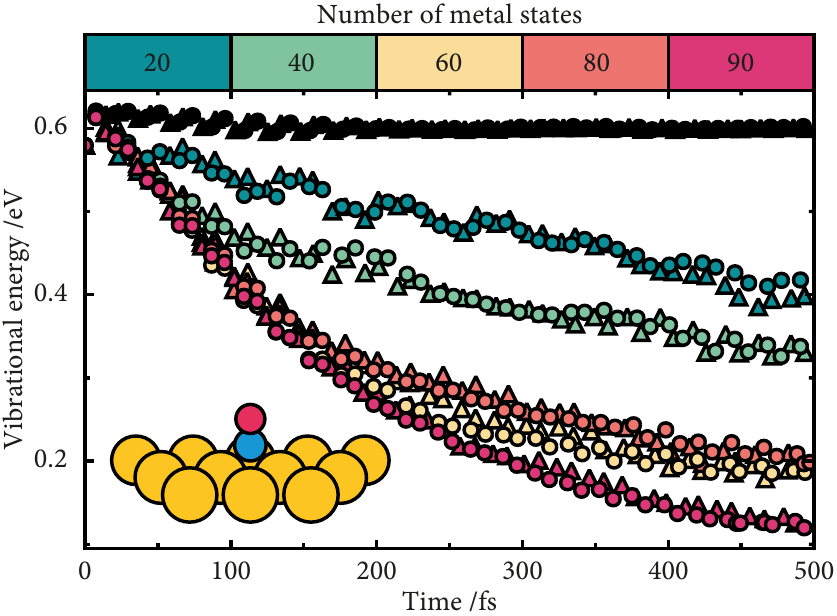}
\caption{
    The vibrational relaxation as a function of time for the NO molecule bound to the Au(111) surface.
    Alongside the results obtained using our NQCDynamics.jl implementation (circles) we show the results obtained using the Fortran program of Ref.~\citenum{krugerNOVibrationalEnergy2015} (triangles).
    The color of the symbols denotes the number of metal states labelled in the color bar.
    The black symbols show the result obtained when performing classical adiabatic dynamics with 40 states.
    A schematic of the atomistic system used for the simulations is shown in the bottom left corner.
}
\label{fig:iesh-tully}
\end{figure}

Each trajectory begins with identical initial conditions.
The NO molecule is positioned vertically above the HCP site with the N atom \SI{1.736}{\angstrom} above the surface,
where the bond length is taken to be the equilibrium gas phase bond length of \SI{1.151}{\angstrom}.
The molecule begins in an excited vibrational state ($\nu_i=2$) with the kinetic component of the vibrational energy equal to \SI{0.58}{\eV}.
The temperature of the system is set to \SI{0}{\kelvin}, such that the Au(111) surface takes its minimum energy configuration and the electronic occupations are initialised with a ground-state Fermi-Dirac distribution.
The Au(111) surface consists of 528 atoms arranged into 4 layers $(11 \times 12 \times 4)$, with the bottom layer frozen throughout the dynamics.
The band width of the metallic bath is \SI{7}{\electronvolt} and we discretize the continuum using the Gauss-Legendre method described in Appendix~\ref{sec:discretization}.
The number of electrons is always equal to half the total number of metal states.
We perform 1000 trajectories with a timestep of \SI{0.1}{\femto\second} and during each trajectory the vibrational kinetic energy of the NO molecule is recorded.
After averaging the kinetic energy profiles obtained from each trajectory, we identify the maximum vibrational kinetic energy during each period and display the result in Fig.~\ref{fig:iesh-tully}.

Initially, the NO on Au(111) simulations were performed to reproduce Fig.~5 of \citet{shenviNonadiabaticDynamicsMetal2009}.
The non-monotonic convergence pattern observed, along with the shape of the relaxation profile, closely matches the original publication but closer inspection reveals
that our implementation consistently overestimates the rate of relaxation.
To further investigate this discrepancy, we repeated the simulations with the Fortran program used in Ref.~\citenum{krugerNOVibrationalEnergy2015}.
The close agreement between these two attempts can be seen in Fig.~\ref{fig:iesh-tully} and further supports the correctness of our implementation.
The origin of the discrepancy between our simulations and the previous results remains unclear, perhaps caused by minor differences in initial conditions.
However, we can be confident that our implementation is correct by demonstrating agreement with the Fortran implementation. 

\subsection{Electronic relaxation in a double-well}\label{sec:double-well}

In addition to the NO on Au(111) model of Sec.~\ref{sec:NOAu},
IESH has been previously applied to a 1D Newns-Anderson Hamiltonian with double-well diabatic surfaces.
This model is useful for modelling electron transfer for which Marcus theory can be used to provide reference results.
\cite{miaoComparisonSurfaceHopping2019}
In this section our implementation is further verified by reproducing the zero temperature IESH results
of Fig.~5c and~5d in Ref.~\citenum{miaoComparisonSurfaceHopping2019}.
These simulations investigate the relaxation of a thermally-excited distribution back to equilibrium,
determining how effectively IESH is able to describe the correct long- and short-time dynamics.

For the double-well model, the two diabatic surfaces are given by
\begin{align}
    U_0(x) &= \frac{1}{2}m\omega^2x^2,\\
    U_1(x) &= \frac{1}{2}m\omega^2(x-g)^2 + \Delta G,
\end{align}
with $m = 2000$, $\omega = \num{2e-4}$, $g = 20.6097$ and $\Delta G = \num{-3.8e-3}$.
The negative sign of $\Delta G$ means that the equilibrium distribution favors population of the $U_1$ state.
All parameters for this model are given in atomic units.
For each trajectory, the initial nuclear degrees of freedom are sampled from the
Boltzmann thermal equilibrium distribution of the neutral $U_0$ state:
\begin{equation}
    \rho(x, p) = \frac{\beta\omega}{2\pi}
    \exp\left[
        -\beta \left(\frac{m\omega^2}{2}x^2 + \frac{p^2}{2m}\right)
    \right].
\end{equation}
To facilitate the non-equilibrium relaxation, the system is prepared at the elevated temperature of $5kT$, such that $\beta = 1/5kT$,
with the equilibrium thermal energy $kT = \num{9.5e-4}$.
Consistent with previous work,\cite{miaoComparisonSurfaceHopping2019} we choose $\Gamma = \num{6.4e-3}$,
the number of bath states as 40, and the bath band width as $10\Gamma$.
The results are averaged over 1000 trajectories with a time step of 100.
The hole-impurity population is calculated as $1 - P_d$, where $P_d$ is the diabatic population of the impurity state calculated using
the method of Ref.~\citenum{pradhanDetailedBalanceIndependent2022}.

\begin{figure}
\includegraphics{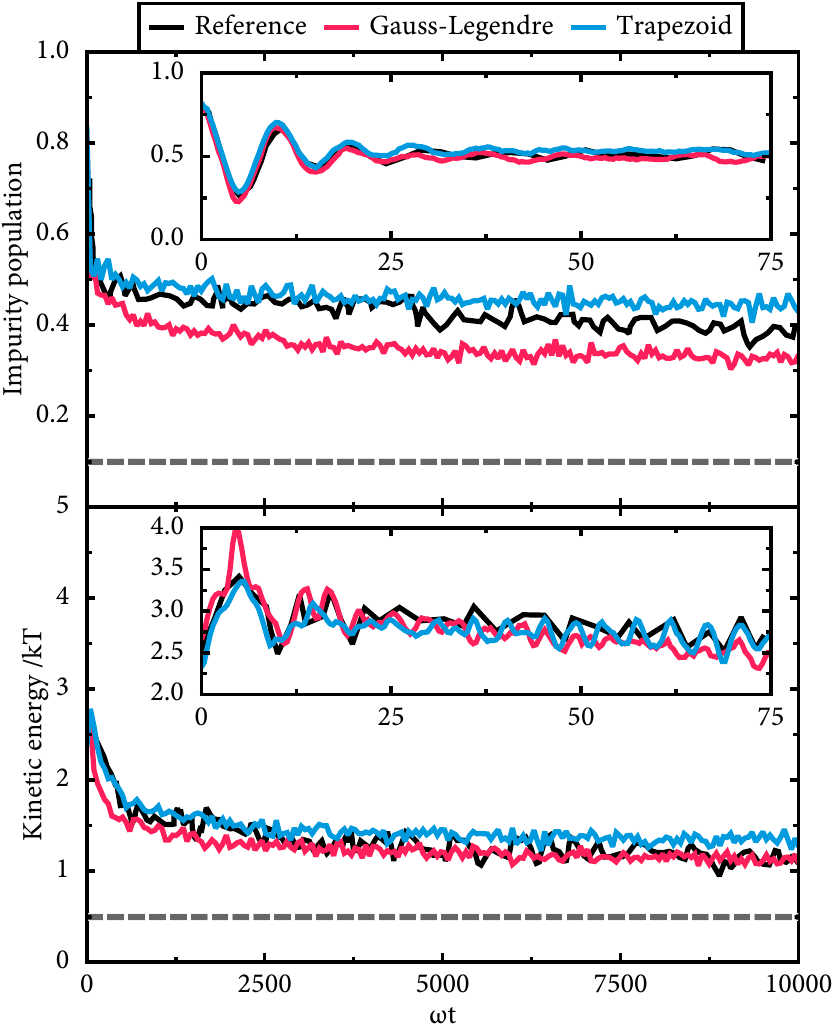}
\caption{
The hole-impurity population (top) and kinetic energy (bottom) during IESH simulations
of the double-well model with the nuclear subsystem prepared at a temperature of $5kT$.
The inset panels show the short time behavior of the same quantities.
Results are shown for both discretization methods.
The reference data is the IESH (zero $T$) data taken from the second column of Fig.~5 in Ref.~\citenum{miaoComparisonSurfaceHopping2019}.
Our plot is formatted similarly such that comparison can be made with the other methods presented in Ref.~\citenum{miaoComparisonSurfaceHopping2019}.
The dashed horizontal lines show the long-time limits for the Marcus theory hole-impurity population and the classical equilibrium kinetic energy.\cite{miaoComparisonSurfaceHopping2019}
}
\label{fig:iesh-subotnik}
\end{figure}

Fig.~\ref{fig:iesh-subotnik} shows the time evolution of the hole-impurity population and kinetic energy
of the particle in the double-well Hamiltonian.
Our simulations have been performed using both discretization methods,
though the reference data\cite{miaoComparisonSurfaceHopping2019} uses only the evenly spaced trapezoid method.
As such, it is expected that the trapezoid result and reference should match exactly.

The short time behavior shown in the insets of Fig.~\ref{fig:iesh-subotnik} display excellent agreement,
regardless of discretization method.
Similarly, the long-time kinetic energy results closely align.
The only discrepancy lies in the long-time population dynamics, where our trapezoid result
slightly overestimates the hole-impurity and the Gauss--Legendre underestimates compared to the IESH reference result.

Included within Fig.~\ref{fig:iesh-subotnik} are the Marcus theory long-time limits\cite{miaoComparisonSurfaceHopping2019} of each quantity.
With these in mind, it appears that IESH can perform better when using the Gauss--Legendre method.
Our results are able to reaffirm the conclusions presented in Ref.~\citenum{miaoComparisonSurfaceHopping2019}
regarding the shortcoming of IESH with respect to predicting the long-time relaxation process.
However, it appears that an enhancement of the relaxation is observed by changing the discretization method.
With sufficiently many states in the bath, both discretization methods are expected to recover the same result.
The disagreement here suggests that the trapezoid result is not converged with only 40 states.
This observation highlights the difficulty of obtaining robust and converged results using IESH. 

\subsection{Molecular desorption in one dimension}\label{sec:desorption}

A model for single-molecule junctions has been previously investigated by Erpenbeck \textit{et al.}
\cite{erpenbeckCurrentInducedBondRupture2018,erpenbeckHierarchicalQuantumMaster2019,erpenbeckCurrentinducedDissociationMolecular2020},
involving a coupling of electronic leads to a molecular subsystem.
When using a single lead, the same model can be used to represent an atom or molecule interacting with a metal surface as in the NA Hamiltonian
of Sec.~\ref{sec:newns-anderson}.
We are interested in modelling the initially neutral molecule located in the minimum of $U_0$, then probing the desorption dynamics of the molecule to assess how quickly it desorbs and the probability with which it desorbs in the long-time limit.

\begin{figure}
    \includegraphics{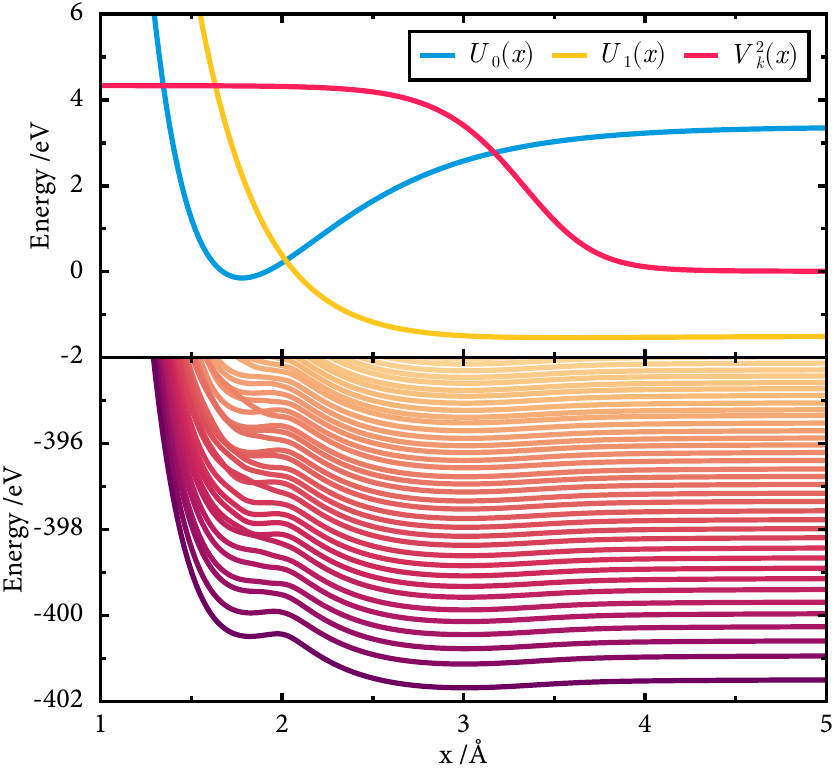}
    \caption{
        The diabatic adsorbate states and coupling given by Eqs.~\ref{eq:U0}--\ref{eq:Vk} (top)
        and the resulting adiabatic surfaces in the NA model used to propagate the nuclei in IESH (bottom).
        For both, $\Gamma = \SI{1}{\eV}$.
        The discretization used to produce the bottom panel is the Gauss-Legendre method using 100 states,
        50 electrons, and a band width of \SI{32}{\eV}.
        A reduced sample of all possible states is shown.
        The lines are colored based on the energy of each state.
    }
    \label{fig:model}
\end{figure}

The model is shown in Fig.~\ref{fig:model}, displaying the two diabatic PESs, the molecule-metal coupling,
and possible adiabatic surfaces that appear in Eq.~\ref{eq:classical_hamiltonian}.
The desorption dynamics are governed by the population transfer from the neutral binding potential to the charged repulsive state.
In the adiabatic picture, a barrier to desorption exists, but can be overcome via electronic excitation to higher energy states with reduced barriers.  

The potential for the neutral molecule is a Morse potential
\begin{equation}
    U_0(x) = D_e\left(
        e^{-a(x-x_0)} - 1
    \right)^2 + c,
    \label{eq:U0}
\end{equation}
but, when charged, the potential becomes repulsive
\begin{equation}
    U_1(x) = D_1e^{-2a'(x-x_0)}
    - D_2e^{-a'(x-x_0)} + V_\infty.
    \label{eq:U1}
\end{equation}
The coupling between the molecule and surface is
\begin{equation}
    V_k(x) = \bar{V}_k \left(
        \frac{1-q}{2}\left[
            1 - \tanh\left(
                \frac{x-\tilde{x}}{\tilde{a}}
            \right) + q
        \right]
    \right).
    \label{eq:Vk}
\end{equation}
Parameters for Eqs.~\ref{eq:U0}--\ref{eq:Vk} are provided in Table~\ref{tab:et-parameters}.
The magnitude of the coupling $\bar{V}_k = \sqrt{\Gamma / 2\pi}$ varies for each set of results.

\begin{table}
    \caption{\label{tab:et-parameters}
    Parameters for the desorption model defined in Eqs.~\ref{eq:U0}--\ref{eq:Vk}.
    }
    \begin{ruledtabular}
    \begin{tabular}{lll|lll|lll}
    $D_e$ & \num{3.52} & \si{\electronvolt} &
    $x_0$ & \num{1.78} & \si{\angstrom} &
    $a$ & \num{1.7361} & \si{\per\angstrom} \\
    $D_1$ & \num{4.52} & \si{\electronvolt} &
    $D_2$ & \num{0.79} & \si{\electronvolt} &
    $a'$ & \num{1.379} & \si{\per\angstrom}\\
    $V_\infty$ & \num{-1.5} & \si{\electronvolt}&
    $q$ & \num{0.05} & 1 &
    $\tilde{a}$ & \num{0.5} & \si{\angstrom}\\
    $\tilde{x}$ & \num{3.5} & \si{\angstrom}&
    $m$ & \num{10.54} & \si{\atomicmassunit} &
    $c$ & \num{-45.7} & \si{\milli\eV} \\
    \end{tabular}
    \end{ruledtabular}
\end{table}

The vibronic model has been used with two different values for the nuclear mass: \SI{1}{\atomicmassunit} and \SI{10.54}{\atomicmassunit},
at a temperature of \SI{300}{\kelvin}.
\cite{erpenbeckCurrentInducedBondRupture2018,erpenbeckHierarchicalQuantumMaster2019}
Using the harmonic frequency of the ground state potential $\omega = \sqrt{\frac{\partial^2 U_0(x)}{\partial x^2}/m}$,
the relative magnitude of $\hbar\omega$ and $kT$ can be compared to estimate the validity of a classical approximation for the nuclear motion.
When $kT > \hbar\omega$ the model is in the regime of classical nuclear motion, but for the parameters previously investigated
$kT / \hbar\omega$ has the value of \SI{0.0868}{} (\SI{1}{\atomicmassunit}) and \SI{0.281}{} (\SI{10.54}{\atomicmassunit}).
In both cases, the nuclear motion is expected to require a quantum description, however, for the larger mass a classical description is more likely to be valid.
Since we are presently interested in the performance of the IESH algorithm for treating nonadiabatic effects in the classical nuclear regime, we will adopt the parametrization with the increased mass of \SI{10.54}{\atomicmassunit}. As quantum nuclear effects may still play a role in this case, some level of caution is advised when comparing with HQME results.

To investigate the desorption probabilities for the vibronic model of \citet{erpenbeckCurrentInducedBondRupture2018} using IESH
we first initialize the nuclear degrees of freedom by sampling the thermal equilibrium distribution of the neutral potential $U_0(x)$ (see Eq.~\ref{eq:U0}).
As in previous work,\cite{erpenbeckCurrentInducedBondRupture2018} we sample the Wigner distribution corresponding to the harmonic approximation to $U_0(x)$,
\begin{multline}
    \rho(x, p) = \frac{\beta\omega}{2\pi Q(\beta,\omega)}
    \\
    \times \exp\left[
        -\frac{\beta}{Q(\beta,\omega)}
        \left(\frac{m\omega^2}{2}(x - x_0)^2 + \frac{p^2}{2m}\right)
    \right],
    \label{eq:wigner}
\end{multline}
where $Q(\beta,\omega) = \beta\hbar\omega/(2\tanh(\beta\hbar\omega/2))$ is the quantum correction to the classical Boltzmann distribution.\cite{liuSimpleModelTreatment2009}
At \SI{300}{\kelvin} with the selected mass of \SI{10.54}{\atomicmassunit},
$Q(\beta,\omega) \approx 1.9$ suggesting the Wigner distribution moderately deviates from the Boltzmann distribution under these conditions.

The electronic degrees of freedom are initialized consistent with a  \SI{300}{\kelvin} Fermi--Dirac distribution.
In all of our simulations, the number of electrons is equal to half the total number of metal states.
The wavefunction coefficients for each electron are initialized to match the discrete populations.

The total desorption probability is given by\cite{erpenbeckCurrentInducedBondRupture2018}
\begin{equation}
    P_\text{total}(t) = \frac{1}{N}\sum_{i=1}^{N} \theta ( x_i(t) - x_\text{threshold})
\end{equation}
where $\theta$ is the Heaviside step function and $x_i(t)$ is the position throughout the trajectory.
As implied by the step function, each trajectory is counted as desorbed when it exceeds $x_\text{threshold}$,
with $x_\text{threshold} = \SI{5}{\angstrom}$ as defined in Ref.~\citenum{erpenbeckCurrentInducedBondRupture2018}.

It has been established that converging IESH simulations, particularly in the wideband limit can be challenging.\cite{miaoComparisonSurfaceHopping2019}
To reach the wideband limit, the band width must be much larger than the metal-molecule coupling $\Gamma$
and, for IESH to probe the electron transfer between molecule and metal, the spacing between the states at the Fermi level must be small compared to $\Gamma$.
In combination, these requirements demand an increase of both the band width and the number of states in order to reach convergence.
Before proceeding to an in-depth analysis of the performance of IESH for this model, it is useful to first investigate the convergence behavior with respect to the number of states and band width.

\begin{figure}
    \includegraphics{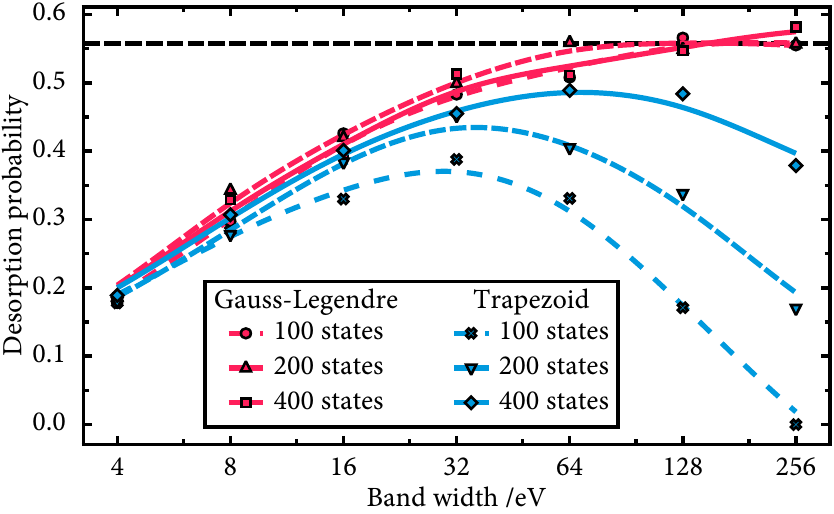}
    \caption{The long-time desorption probability for the vibronic model as a function of the band width with $\Gamma = \SI{1}{\eV}$.
    Results are shown for the two discretization methods introduced in Sec.~\ref{sec:newns-anderson}.
    For each discretization method, 100 (dotted), 200 (dashed) and 400 (solid) metal states have been used.
    The dashed horizontal line displays the result obtained in Ref.~\citenum{erpenbeckCurrentInducedBondRupture2018}.
    }
    \label{fig:convergence}
\end{figure}

Fig.~\ref{fig:convergence} shows the long-time desorption probabilities obtained when $\Gamma = \SI{1}{\eV}$.
For small band widths (\SIrange{4}{16}{\eV}) both discretization methods display equivalent convergence behavior,
however, for larger band widths (\SIrange{32}{256}{\eV}) the Gauss--Legendre method approaches the converged result
and the trapezoid method falls away, underestimating the desorption probability.
This effect becomes more pronounced when using fewer states.
The Gauss--Legendre method shows minimal dependence on the number of states we have investigated, but the trapezoid method fails to reach convergence even with 400 states.
Clearly, for this choice of $\Gamma$, the Gauss--Legendre discretization method outperforms the simpler trapezoid method.
In subsequent simulations we have used the converged settings of 100 states and a band width of \SI{64}{\eV} with the Gauss--Legendre discretization.
We have additionally checked the convergence for different values of $\Gamma$ and observe that the chosen settings are  valid for all relevant values of $\Gamma$.

\begin{figure}
    \includegraphics{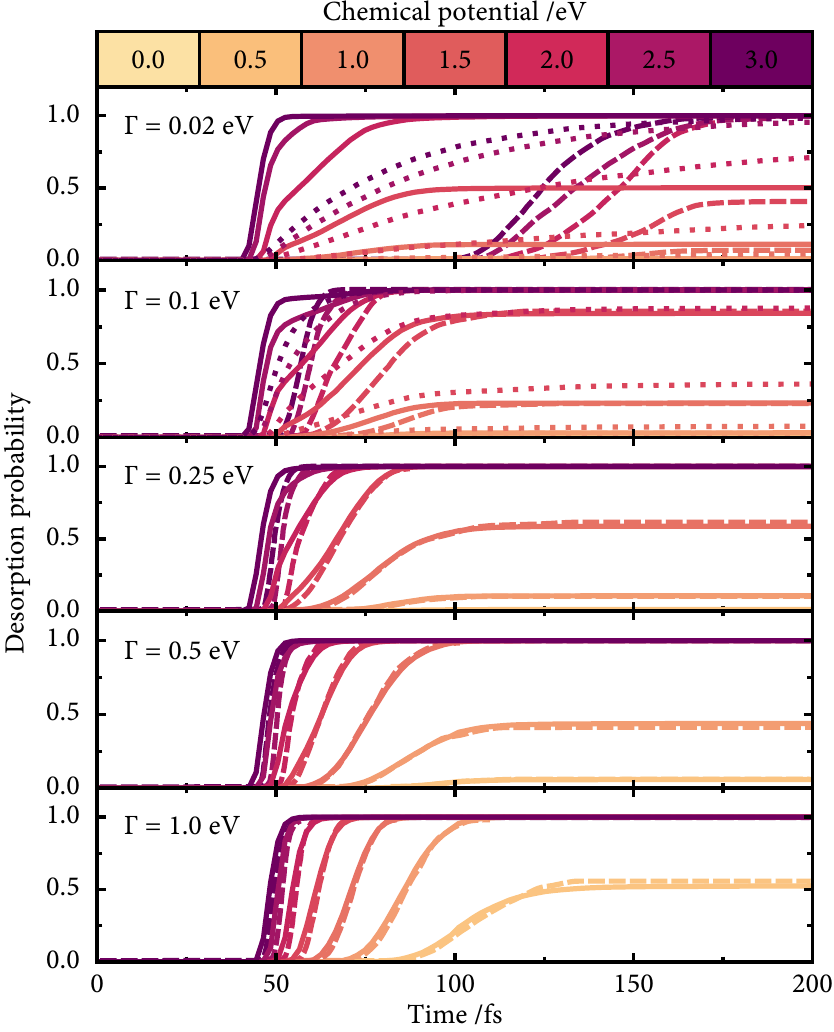}
    \caption{Desorption probability as a function of time for the 1D model of \citet{erpenbeckCurrentInducedBondRupture2018}.
    The solid lines show the result of the IESH simulations, the dashed lines are the results taken from Ref.~\citenum{erpenbeckCurrentInducedBondRupture2018}
    using the HQME method in combination with Ehrenfest for the nuclear degrees of freedom.
    From top to bottom, each panel has increasing molecule-metal coupling $\Gamma$ as labelled.
    \protect\highlightred{Results obtained using the CME method are shown for the two smallest values of $\Gamma$ using dotted lines.}
    Each line corresponds to increasing chemical potential as indicated in the color bar.
    }
    \label{fig:desorption}
\end{figure}

\begin{figure}
    \includegraphics{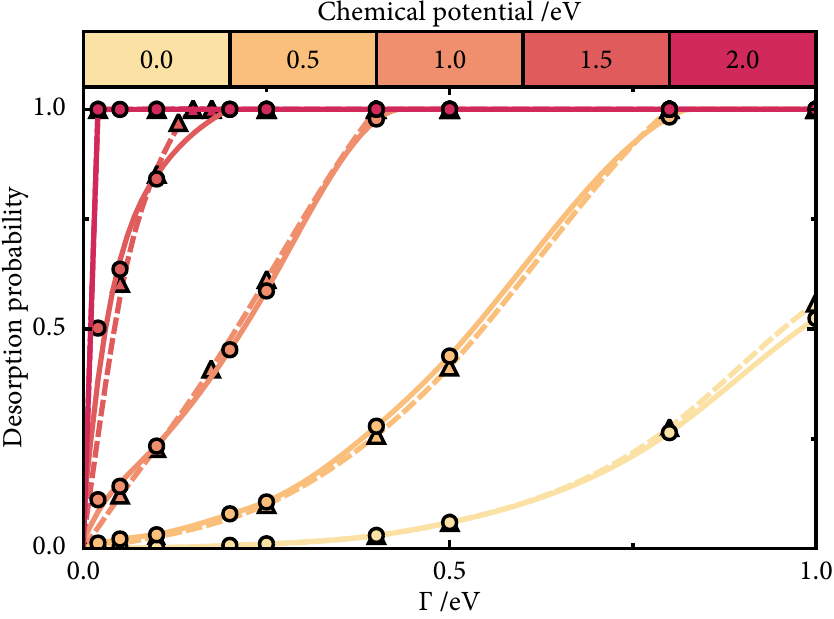}
    \caption{The long-time desorption probabilities as a function of molecule-metal coupling for increasing chemical potentials shown in the color bar.
    The data points in this figure correspond directly to the \SI{200}{\femto\second} probabilities in Fig.~\ref{fig:desorption}.
    Additional intermediate values for $\Gamma$ were added to better characterise the desorption profile.
    The reference data from Ref.~\citenum{erpenbeckCurrentInducedBondRupture2018} is shown with triangles and dashed lines,
    the IESH results are shown with circles and solid lines.
    }
    \label{fig:desorption_longtime}
\end{figure}

\begin{figure}
    \includegraphics{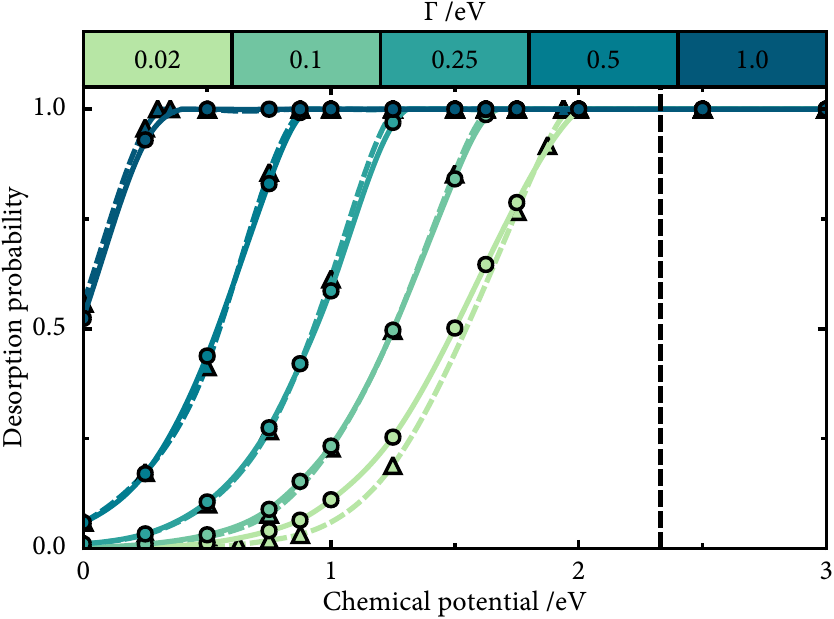}
    \caption{The long-time desorption probabilities as a function of chemical potential for increasing molecule metal coupling shown in the color bar.
    The data points in this figure correspond directly to the \SI{200}{\femto\second} probabilities in Fig.~\ref{fig:desorption}.
    Additional results with intermediate chemical potentials were added to better characterise the desorption profile.
    The reference data from Ref.~\citenum{erpenbeckCurrentInducedBondRupture2018} is shown with triangles and dashed lines,
    the IESH results are shown with circles and solid lines.
    The vertical dashed line denotes the chemical potential at which the dissociation becomes barrierless.
    }
    \label{fig:desorption_longtime_mu}
\end{figure}

Having established convergence, we can now proceed to investigate the full desorption dynamics.
Fig.~\ref{fig:desorption} displays the desorption probablity as a function of time for different values of $\Gamma$ and chemical potential.
Alongside our IESH results, we have also included the mixed-quantum classical HQME results
of \citet{erpenbeckCurrentInducedBondRupture2018} for reference.
The HQME method employs a numerically exact treatment of the electronic subsystem,
treating the system-bath interaction as an open quantum system.
However, using the Ehrenfest approach to model the electron-nuclear coupling is an approximation
that fails to describe some physical effects such as Joule heating.
\cite{horsfieldPowerDissipationNanoscale2004,horsfieldEhrenfestCorrelatedNonadiabatic2004}
Despite this approximation, the exact treatment of the system-bath coupling should lead to a more accurate description of the desorption process.
As such, we deem IESH to be performing well when it closely matches the reference result.
\highlightred{
However, when $\Gamma$ is small, Ehrenfest becomes a less reliable reference.
To provide a better benchmark in this regime, we use the classical master equation (CME) method,\cite{douSurfaceHoppingManifold2015a} 
another variant of surface hopping, that performs well when $\Gamma$ is small.
However, we also note that CME assumes the Condon approximation, where $\Gamma$ is independent of position.
}
Figs.~\ref{fig:desorption_longtime} and~\ref{fig:desorption_longtime_mu} display the long-time probabilities
extracted from Fig.~\ref{fig:desorption} to highlight the trends as a function of $\Gamma$ and chemical potential.
A comprehensive discussion of the physical phenomena demonstrated by this model has been provided previously by
\citet{erpenbeckCurrentInducedBondRupture2018}.
Here, we will focus on the performance of the IESH algorithm and its ability to capture the dynamics of the model.

First, it is interesting to consider the short-time desorption dynamics, specifically the time at which the molecule begins to desorb.
The short-time behavior is characterized by the onset of the desorption in Fig.~\ref{fig:desorption}.
For strong molecule-metal coupling, IESH performs well, accurately capturing the desorption onset when $\Gamma > \SI{0.25}{\eV}$.
However, upon reduction of $\Gamma$ below \SI{0.25}{\eV}, IESH predicts faster desorption than \highlightred{both CME and HQME}.
For IESH, it appears that the initial onset of the desorption curve is largely independent of $\Gamma$.
Increasing $\Gamma$ increases the rate of desorption but not the time at which desorption begins.
For $\Gamma = \SI{0.02}{\eV}$, this leads to a strong underestimation of the desorption onset time \highlightred{compared to HQME}.
\highlightred{
However, the desorption onset for CME remains at approximately \SI{50}{\femto\second} similarly to IESH and the desorption profile shows a timescale of desorption for CME that is between the IESH and HQME result.
}
The ability of IESH to capture the long-time dynamics can be observed in Figs.~\ref{fig:desorption_longtime} and~\ref{fig:desorption_longtime_mu}.
The agreement is excellent across a wide range of chemical potentials and $\Gamma$ values, with only minor deviations apparent in Fig.~\ref{fig:desorption_longtime_mu} when $\Gamma = \SI{0.02}{\eV}$.

In the case of small $\Gamma$, \highlightred{all three methods show different rates of desorption and} it is reasonable to conclude that IESH cannot accurately describe the short-time dynamics \strike{and shows some deviations from HQME in the long-time regime}.
We note that during the simulations, hopping events are rare, and the results obtained do not significantly deviate from  classical adiabatic dynamics. This suggests that the desorption dynamics are not strongly influenced by nonadiabatic transitions, but rather by the change of the ground-state energy landscape as a function of $\Gamma$ and chemical potential. 
The adiabatic nature of the dynamics can be explained by the low particle velocities encountered under thermal conditions.
\strike{It is therefore possible that the failure of IESH to capture the short timescale desorption dynamics at low $\Gamma$ is due to the neglect of quantum nuclear effects.}
In Sec.~\ref{sec:scattering}, we go beyond thermal conditions and explore the performance of IESH for higher energy scattering events where nonadiabatic effects are significant.

\subsection{Molecular scattering in one dimension}\label{sec:scattering}

Previous simulations using IESH have focused on the scattering of NO on the Au(111) surface\cite{shenviNonadiabaticDynamicsMetal2009,shenviDynamicalSteeringElectronic2009,shenviNonadiabaticDynamicsMetal2012,krugerNOVibrationalEnergy2015}
but IESH has not been applied to scattering in a low-dimensional model where the coupling strength can be modulated to explore different scattering regimes.
The model presented in Sec.~\ref{sec:desorption} can be used to simulate the scattering of a molecule on a metal surface by modifying the initial position and momenta of the molecule. 
In a one dimensional system, fully adiabatic molecular dynamics simulations would predict a perfectly elastic collision, where the final kinetic energy after collision matches the initial.
However, nonadiabatic energy transfer into EHP excitations due to the collision with the surface will lead to inelastic energy loss.
By measuring the change in kinetic energy due to the scattering event, we gain insights into the coupled electron-nuclear dynamics and the ability of IESH to describe the process.

We initialise the adsorbate at a distance of \SI{5}{\angstrom} from the surface and the velocity towards the surface is set to correspond to a given kinetic energy.
The electronic subsystem is initialized at \SI{0}{\kelvin}.
We allow each trajectory to run for \SI{300}{\femto\second}, but terminate early once  the molecule has rebounded to its initial height.
The reported results include only the trajectories that have rebounded from surface before the time limit is reached.
By filtering the data, we neglect a small portion of trajectories that remain trapped on the surface.
For most ensembles, all trajectories rebound before the time limit is reached and the total amount of excluded trajectories is negligible.

In addition to the IESH dynamics, for this scenario we also perform MDEF simulations.
The MDEF equations of motion used for our simulations are
\begin{align}
    \dot{x} &= \frac{p}{m} \\
    \dot{p} &= -\frac{\partial}{\partial x} U_0 - \sum_k f(\lambda_k) \frac{\partial}{\partial x} \lambda_k - \gamma(x)\frac{p}{m} \label{eq:mdef-force}
\end{align}
where the random force does not appear because the electronic subsystem is at a temperature of \SI{0}{\kelvin}.
The first two terms on the right hand side of Eq.~\ref{eq:mdef-force} describe the adiabatic PES.
The Fermi function $f(\lambda_k)$ at \SI{0}{\kelvin} enforces that only the states up to the Fermi level contribute to the force.
The final term describes the dissipation of energy due to excitation of electron-hole pairs in the electronic bath.

The friction coefficient is calculated using the exact expression in the wideband limit\cite{jinPracticalAnsatzEvaluating2019,brandbygeElectronicallyDrivenAdsorbate1995}
\begin{equation}
    \gamma(x) = -\pi\hbar\int \dif\epsilon
    \left(
        \frac{\partial h}{\partial x} + \frac{\epsilon - h}{\Gamma}\frac{\partial \Gamma}{\partial x}
    \right)^2
    A^2(\epsilon)
    \frac{\partial f}{\partial \epsilon}
    \label{eq:exact-friction}
\end{equation}
where
\begin{equation}
    A(\epsilon) = \frac{1}{\pi}\frac{\Gamma/2}{(\epsilon-h)^2 + (\Gamma/2)^2}.
\end{equation}
and recalling from Eq.~\ref{eq:electronic_hamiltonian} that $h = h(x) = U_1(x) - U_0(x)$.
We have additionally confirmed that the Gaussian broadening, off-diagonal normalized Gaussian broadening, and direct quadrature methods
highlighted by \citet{jinPracticalAnsatzEvaluating2019} converge to the same result, but use the exact expression for simplicity.
To evaluate Eq.~\ref{eq:exact-friction} we have used adaptive Gauss--Kronrod quadrature\cite{quadgk}.
During the evaluation, the temperature used to calculate the derivative of the Fermi function $\frac{\partial f}{\partial \epsilon}$ is set to \SI{100}{\kelvin}
to bypass the singularity encountered at \SI{0}{\kelvin}.
In doing so, we have verified that the value of the friction coefficient is largely independent of temperature below room temperature.

\begin{figure}
    \includegraphics{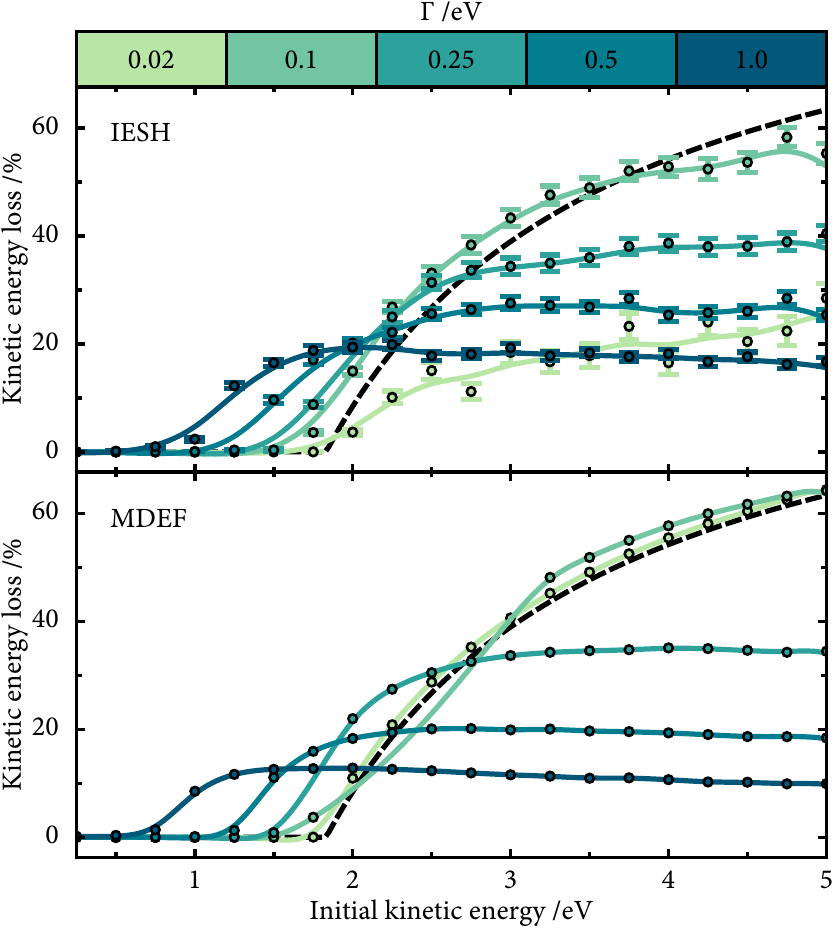}
    \caption{Kinetic energy loss as a function of initial kinetic energy experienced during molecular scattering for the vibronic model introduced by \citet{erpenbeckCurrentInducedBondRupture2018}.
    Results are shown for IESH (top) and MDEF (bottom).
    The circles denote the computed data points and the lines are generated via kernel ridge regression.
    The error bars on each IESH result show the standard error of the mean.
    The dashed black line shows the effective energy loss observed if the particle returns with \SI{1.83}{\eV} kinetic energy.
    \SI{1.83}{\eV} is the height of the potential energy barrier encountered at the crossing point of $U_0$ and $U_1$ as $\Gamma$ tends to zero. 
    }
    \label{fig:scattering}
\end{figure}

In Fig.~\ref{fig:scattering}, we present the kinetic energy loss observed after scattering has occurred.
The results of the IESH simulations in the top panel can be directly compared to the MDEF results in the bottom panel.
First, it is interesting to consider the qualitative trends that appear in both IESH and MDEF simulations.
With incident energy below \SI{0.5}{\eV}, both methods predict adiabatic dynamics, where the collision is completely elastic.
However, upon increasing the kinetic energy from \SI{0.5}{\eV} to \SI{3}{\eV}, all values of $\Gamma$ display a monotonic increase in energy lost to the electronic subsystem.
The initial onset of the monotonic increase depends to the value of $\Gamma$, with larger values leading to an onset at lower incidence energies.
For all values of $\Gamma$ greater than \SI{0.02}{\electronvolt}, the shape of the curves is similar between the two methods.
After the region of initial increase, the energy loss as a function of incidence energy appears to plateau for larger values of $\Gamma$. For smaller values of $\Gamma$, the energy loss increases monotonically across the entire range of kinetic energies.

The most significant difference between the two methods is the $\Gamma = \SI{0.02}{\eV}$ result.
MDEF describes an energy loss profile that closely follows the $\Gamma = \SI{0.1}{\eV}$ result and the dashed black line, whereas IESH predicts a smaller energy loss for $\Gamma = \SI{0.02}{\eV}$ compared to $\Gamma = \SI{0.1}{\eV}$.
The dashed black line shows the relative kinetic energy loss that corresponds to the particle returning with an energy of exactly \SI{1.83}{\eV}. This would be the case if the particle were to stop at the crossing point between the two diabatic states, lose all kinetic energy, and ``roll back down the hill'' to the starting point.
The MDEF result follows the dashed line when $\Gamma$ is small because there is a significant spike in the friction coefficient at the crossing point.
The spike leads to the particle losing all of its initial energy at the top of the barrier, then rolling back down, gaining energy equal to the barrier height.

In contrast, IESH deviates significantly from this behavior, displaying far less energy loss.
Closer inspection of individual trajectories shows that electrons that were excited upon impact with the surface, relax as the molecule leaves the surface, transferring energy back into adsorbate kinetic energy. Therefore, the IESH dynamics have the ability to describe transient excitation and deexcitation of electrons leading to energy transfer in both directions, despite the fact that the initial electronic system was at \SI{0}{\K}.  However, in the MDEF simulations the frictional force allows the molecule only to lose energy to the electronic bath and no random force is observed for a bath at \SI{0}{\K}.
As such, MDEF is unable to describe the bi-directional energy transfer responsible for the IESH $\Gamma = \SI{0.02}{\eV}$ result.

To further investigate the energy transfer, the simulations were repeated with the electronic temperature set to \SI{300}{\K} for both IESH and MDEF.
In both cases, we observe little deviation from the \SI{0}{\K} result presented in Fig.~\ref{fig:scattering}.
Although setting the electronic temperature to \SI{300}{\K} removes the uni-directional energy transfer restriction present in the \SI{0}{\K} MDEF case,
it is clear that MDEF still does not recover the IESH result.
\highlightred{This failure is likely due to inability of MDEF to correctly describe the dynamics when $\Gamma$ is small.\cite{douFrictionalEffectsMetal2015,douNonadiabaticMolecularDynamics2020}
}

\strike{
This can be explained by considering the disparity in the magnitudes of the incident energy and the thermal energy at 300 K.
The random force in MDEF acts to restore thermal equilibrium between the nuclear and electronic subsystems, but in this case,
the incident energy is significantly larger than the thermal energy and the transfer of energy strongly favors transfer into the electronic bath.
}

\section{Conclusions}\label{sec:conclusion}

We have presented an efficient open-source implementation of the IESH method, capable of treating simple model Hamiltonians and full dimensional atomistic systems.
Its correctness has been verified against previously published results,
including vibrational relaxation using a high-dimensional NO on Au(111) Hamiltonian and electronic relaxation in a one-dimensional double-well model.
The agreement observed is good, we find only minor deviations in convergence behavior.
Furthermore, we have used our implementation to investigate a molecular desorption model,
determining its effectiveness in modelling nonadiabatic desorption from a metal surface.
For the desorption model, we have extended the IESH algorithm to treat systems under the influence of an external bias by introducing a shift in the chemical potential.
With this modification, IESH appears capable of accurately capturing the dynamics in biased systems in certain regimes of molecule-metal coupling, however, in the studied regime, the desorption probability does not seem to be strongly affected by EHP excitations.
Finally, we performed scattering simulations with IESH and compared the result to MDEF,
where both methods gave similar results, except when the molecule-metal coupling was small, where MDEF predicted less energy transfer back from the electronic system into the adsorbate.
Further comparison of IESH and MDEF on realistic systems would be useful to affirm this observation.

Our implementation currently matches the original specification of the algorithm\cite{shenviNonadiabaticDynamicsMetal2009}
but further development and algorithmic enhancement is possible.
Our implementation could be expanded to include existing developments such as
decoherence corrections\cite{pradhanDetailedBalanceIndependent2022} and phononic and electronic thermostats\cite{shenviNonadiabaticDynamicsMetal2012},
or extended to include nuclear quantum effects.\cite{shushkovRingPolymerMolecular2012,shakibRingPolymerSurface2017}
To improve computational efficiency, it may be possible to adapt the hopping scheme to allow for larger nuclear time steps,
include adaptive time stepping based on the hopping probabilities, or devise new, more efficient, discretization schemes.
Another avenue to consider is the extension of IESH to model realistic metallic continuums based on \textit{ab initio} band structure calculations.

We believe IESH is a promising method for going beyond MDEF, explicitly including electron dynamics.
It is our hope that the open-source nature and availability of our implementation within NQCDynamics.jl will encourage further use and development of the IESH method,
expanding the toolbox of methods available for nonadiabatic simulation of molecules interacting with metal surfaces.

\begin{acknowledgments}
This work was financially supported by the Leverhulme Trust (RPG-2019-078), the UKRI Future Leaders Fellowship program (MR/S016023/1) (R.J.M.) and the WIRL-COFUND fellowship scheme at the University of Warwick (S.M.J.), under the Marie Skłodowska Curie Actions COFUND program (grant agreement number 713548). High performance computing resources were provided via the Scientific Computing Research Technology Platform of the University of Warwick, the EPSRC-funded Materials Chemistry Consortium (EP/R029431/1) for the ARCHER2 UK National Supercomputing Service,
and the EPSRC-funded HPC Midlands$^+$ computing center for the Sulis service (EP/T022108/1). We thank Dr. Alexander Kandratsenka and Belal Raza (MPI NAT, Göttingen) for fruitful discussions and for providing us with the original Fortran IESH code.
\end{acknowledgments}

\section*{Author declarations}
\subsection*{Conflict of interest}
The authors have no conflicts to disclose.

\subsection*{Author contributions}

\textbf{James Gardner:}
Conceptualization (equal);
data curation;
investigation (lead);
software (lead);
validation;
visualization;
writing -- original draft;
writing -- review and editing (equal).
\textbf{Daniel Corken:} 
Conceptualization (supporting);
investigation (supporting);
software (supporting);
writing -- review and editing (supporting).
\textbf{Svenja M. Janke:}
Investigation (supporting);
software (supporting).
\textbf{Scott Habershon:}
Conceptualization (supporting);
supervision (supporting);
writing -- review and editing (equal).
\textbf{Reinhard J. Maurer:}
Conceptualization (equal);
Investigation (equal);
supervision (lead);
writing -- review and editing (equal).

\section*{Data availability statement}
NQCDynamics.jl is open-source and available at: https://github.com/NQCD/NQCDynamics.jl.
Scripts for generating the data and plotting all of the figures are available at: https://doi.org/10.5281/zenodo.7347918.
Data from Figs.~1-3 and~5-9 are available at wrap.warwick.ac.uk/170983.

\appendix

\section{Discretizing the metallic continuum}\label{sec:discretization}

As described in Refs.~\citenum{shenviEfficientDiscretizationContinuum2008} and~\citenum{devegaHowDiscretizeQuantum2015},
the procedure to obtain $\{\epsilon_k\}$ and $\{V_k\}$ in Eq.~\ref{eq:discrete_electronic_hamiltonian} involves a discretization of the coupling integral that appears in the retarded system Green's function.
This continuous form is
\begin{equation}
    \Lambda(E) = \int_{a}^b \dif\epsilon \, \frac{|V(\epsilon)|^2}{E - \epsilon},
\end{equation}
where the simplest method to discretize the integral uses the trapezoid rule to give
\begin{equation}
    \Lambda(E) = \sum_{k=1}^N \frac{|V(\epsilon_k)|^2}{E - \epsilon_k} \frac{b-a}{N}.
\end{equation}
Inspection of the discretized form allows us to identify 
the coupling values in Eq.~\ref{eq:discrete_electronic_hamiltonian} as $V_k = V(\epsilon_k) \sqrt{(b-a)/N}$
with corresponding energy values $\epsilon_k = a + (k-1)(b-a)/N$.

The Gaussian quadrature approach of \citet{shenviNonadiabaticDynamicsMetal2009} uses Gauss-Legendre
quadrature to discretize this integral in two halves, split at the Fermi level,
\begin{equation}
    \Lambda(E)
    = \int_{a}^{\epsilon_f} \dif \epsilon \, \frac{|V(\epsilon)|^2}{E - \epsilon}
    + \int_{\epsilon_f}^{b} \dif \epsilon \, \frac{|V(\epsilon)|^2}{E - \epsilon}.
    \label{eq:gaussian-quadrature}
\end{equation}
Using Gauss-Legendre quadrature, this becomes
\begin{equation}
    \Lambda(E)
    = \sum_{k=1}^N \frac{|V(\epsilon_k^\text{lower})|^2}{E - \epsilon_k^\text{lower}} w_k^\text{lower}
    + \sum_{k=1}^N \frac{|V(\epsilon_k^\text{upper})|^2}{E - \epsilon_k^\text{upper}} w_k^\text{upper}
\end{equation}
where the knots $\epsilon_k^\text{upper/lower}$ and weights $w_k^\text{upper/lower}$ are obtained via a rescaling of the Gauss--Legendre knots $x_k$ and weights $w_k$ obtained using standard algorithms,\cite{golubCalculationGaussQuadrature1969}
\begin{align}
    \epsilon_k^\text{lower} &= \frac{\epsilon_f-a}{2}x_k + \frac{a+\epsilon_f}{2}
    \\
    \epsilon_k^\text{upper} &= \frac{b - \epsilon_f}{2}x_k + \frac{\epsilon_f+b}{2}
    \\
    w_k^\text{lower} &= \frac{\epsilon_f-a}{2} w_k
    \\
    w_k^\text{upper} &= \frac{b - \epsilon_f}{2} w_k.
\end{align}
With this, the coupling values can be identified as $V_k = V(\epsilon_k)\sqrt{w_k^\text{upper/lower}}$,
and the bath energies as $\epsilon_k = \epsilon_k^\text{upper/lower}$.
Setting $\epsilon_f = 0$, and placing $\epsilon_f$ at the center of the band,
it is possible to simplify the expressions above to obtain
\begin{align}
    \epsilon_k^\text{lower} &= \frac{\Delta E}{4}x_k - \frac{\Delta E}{4}
    \\
    \epsilon_k^\text{upper} &= \frac{\Delta E}{4}x_k + \frac{\Delta E}{4}
    \\
    w_k^\text{lower} &= -\frac{\Delta E}{4} w_k
    \\
    w_k^\text{upper} &= \frac{\Delta E}{4} w_k.
\end{align}
These are consistent with the values presented by \citet{shenviNonadiabaticDynamicsMetal2009}.
The two discretization strategies described in this section are the only choices that have previously been used for IESH simulations.
\cite{miaoComparisonSurfaceHopping2019, shenviNonadiabaticDynamicsMetal2009}

\section{Mapping \textit{ab initio} Hamiltonians onto the Newns-Anderson model}\label{sec:mapping-hamiltonian}

For molecule-metal systems such as NO on Au,
it is possible to generate two state diabatic Hamiltonians
where the two states represent the neutral and anionic states of the molecule.
\textit{Ab initio} methods can provide the adiabatic ground state and different excited-state methods and diabatization procedures can provide the diabatic two-state Hamiltonian.
\strike{The eigenvalues of the two state Hamiltonian correspond directly to the adiabatic energies.}
\highlightred{The adiabatic and diabatic Hamiltonians are related by a similarity transformation:}
\begin{equation}
H_\text{adiabatic} = 
\begin{pmatrix}
E_0 & 0 \\
0 & E_1 \\
\end{pmatrix}
=
\mathbf{P}^T
\begin{pmatrix}
U_0 & C \\
C & U_1
\end{pmatrix}
\mathbf{P}.
\end{equation}

To map the two state model onto the NA model with many electronic states, it is necessary to ensure
the same ground state adiabatic energy \highlightred{$E_0$} is returned by both the two state model and the full NA model \highlightred{(Eq.~\ref{eq:discrete_electronic_hamiltonian})}.
This is the procedure briefly described by \citet{shenviNonadiabaticDynamicsMetal2009} where the two-state model
parameters are adjusted to ensure the ground state energy is conserved.

\highlightred{
Recall that the discrete form of the electronic NA Hamiltonian can be written as in Eq.~\ref{eq:discrete_electronic_hamiltonian},
where $h = U_1 - U_0$.
As explained in Appendix~\ref{sec:discretization}, the NA coupling elements $V_k$ are obtained from the coupling function $V(\epsilon)$ associated with the
hybridization function (Eqs.~\ref{eq:hybridization} and~\ref{eq:discrete_hybridization}) via discretization.
However, in the case of the two-state diabatic model, the hybridization function is not available.
Instead, the discrete coupling values $V_k$ must be chosen such that the NA ground state matches the true ground state energy:
\begin{equation}
    E_0 = U_0 + \sum_{i} f(\lambda_i) \lambda_i,
    \label{eq:adiabatic_energy}
\end{equation}
where $\{\lambda_i\}$ are the eigenvalues of the NA hamiltonian and $f(\lambda_i)$ is the Fermi function that determines the ground-state electronic occupations.
}

\highlightred{
A simple procedure to obtain the coupling elements works as follows.
First, a trial expression for $V_k$ is defined
\begin{equation}
V_k(x) = a(x) C(x) \sqrt{w_k}
\label{eq:trial}
\end{equation}
where $C(x)$ is the diabatic coupling, $w_k$ is a discretization dependent weight ($(b-a)/N$ or obtained from the Gauss-Legendre weights (see Appendix~\ref{sec:discretization}), and $a(x)$ is a fitting parameter.
For each position $x$, the value of $a(x)$ can be optimised to satisfy Eq.~\ref{eq:adiabatic_energy}.
By using Eq.~\ref{eq:trial}, $a(x)$ becomes only a small correction to the diabatic coupling $C(x)$.
In this procedure, the product $a(x)C(x)$ approximates the coupling function in Eqs.~\ref{eq:hybridization} and~\ref{eq:discrete_hybridization}
and must have units of $(\text{energy})^{-1/2}$.
In the case of the NO on Au model in Sec.~\ref{sec:NOAu}, $a(x)$ is independent of position,
and becomes a global scaling factor for the diabatic coupling.
}

\strike{
Consider the two state adiabatic Hamiltonian obtained from \textit{ab initio} electronic structure calculations
containing the energies of the adiabatic states on the diagonal:
If the two diabatic states are known via some diabatization procedure then the related diabatic Hamiltonian
can be set up by selecting the off-diagonal coupling elements $C$ such that diagonalization of the diabatic
Hamiltonian gives the adiabatic form
}

\strike{
The Newns-Anderson Hamiltonian is constructed from the diabatic Hamiltonian by placing the difference between
the diabatic states onto the diagonal, then filling the rest with the energy of the bath states.
The form of the coupling elements is unknown, though they are related to the diabatic coupling $C$
}

\strike{
The value of $c$ is constrained such that the eigenvalues $\{\lambda_i\}$ of $H_\text{NA}$ satisfy the following relationship
which states that the adiabatic ground state of the original Hamiltonian $E_0$ is equal to the sum of energy of the first diabatic state
and all the occupied single electron states in $H_\text{NA}$.
This ensures that the ground state energies of both models are the same and classical adiabatic dynamics will recover the same result.
}

\bibliography{IESH}

\end{document}